# Global phase diagram for the quantum Hall effect: An experimental picture

S. V. Kravchenko, Whitney Mason, and J. E. Furneaux

*Laboratory for Electronic Properties of Materials and Department of Physics and Astronomy, University of Oklahoma,*
*Norman, Oklahoma 73019*

V. M. Pudalov

*Institute for High Pressure Physics, Troitsk, 142092 Moscow District, Russia*

(June 10, 1995)

We study the behavior of the extended states of a two-dimensional electron system in silicon in a magnetic field, $B$. Our results show that the extended states, corresponding to the centers of different Landau levels, *merge* with the lowest extended state as $B \to 0$. Using our data, we construct an experimental-based "disorder *vs* filling factor" phase diagram for the integer quantum Hall effect (QHE). Generalizing this diagram to the case of the fractional QHE, we show that it is consistent with the recently observed direct transitions between insulator and fractional QHE at $\nu = 2/5$, $2/7$, and $2/9$.

It is generally accepted that in a two-dimensional electron system (2DES), placed in a strong perpendicular magnetic field, $B$, there exist extended states at the center of each Landau level [1]. In a decreasing magnetic field, the extended states belonging to different levels are predicted neither to vanish nor to merge, but to float up in energy and finally leave the Fermi sea [2,3] leading to an insulating state at $B = 0$. In spite of the fundamental importance of this prediction, there were no attempts to test it experimentally for almost 10 years. The interest in this and related problems was recently revived when magnetic-field-induced transitions between insulating and quantum Hall states were observed almost simultaneously in GaAs/AlGaAs heterostructures [4–6] and Si metal-oxide-semiconductor field-effect transistors (MOSFET's) [7–9]. In addition, theoretical global phase diagrams for the quantum Hall effect (QHE), recently proposed by Kivelson, Lee, and Zhang (KLZ) [10] by Halperin, Lee, and Read (HLR) [11], have fueled experimental studies of the behavior of extended states in disordered two-dimensional systems.

Shashkin, Kravchenko, and Dolgopolov (SKD), studying a 2DES in silicon, reported the floating up of extended states with respect to their expected positions at half-integer filling factors $\nu \equiv n_s hc/eB = i + \frac{1}{2}$ as $B \to 0$ [12,13] (here $n_s$ is the electron density, and $i$ is the level index). However, contrary to the theoretical predictions of Khmelnitskii [2], Laughlin [3], and KLZ, the extended states did not float up indefinitely; instead, they coalesced at some finite energy as $B \to 0$. To map positions of the extended states, SKD used two methods. For the first method, the system was considered metallic if the diagonal conductivity, $\sigma_{xx}$, was higher than $e^2/20h$ and insulating otherwise. For the second, the metal/insulator boundary was determined using the criterion of vanishing activation energy [9]. Both methods gave similar results.

Later, Glozman, Johnson, and Jiang (GJJ) [14] studied a low-mobility gated GaAs/AlGaAs structure and reported floating up of extended states *without merging* in agreement with theoretical predictions. They mapped the positions of extended states by identifying them as peaks in $\sigma_{xx}$. The principle discrepancy of their results with those of Ref. [12] was attributed to some arbitrariness in the first of two criteria used by SKD.

To resolve this inconsistency, we have performed a systematic study of the position of extended states in very high-quality silicon MOSFET's using the method suggested by GJJ. We have found that the extended states belonging to different Landau levels *do merge* with the lowest one in a vanishing magnetic field, while the levels remain well-resolved to the point of merging. Using the obtained map of the extended states, we construct an experimental-based "disorder *vs* filling factor" phase diagram for the integer QHE. Applying correspondence principles articulated by KLZ, we construct a global phase diagram which appears to be consistent with direct transitions between an insulating state and the fractional QHE (FQHE) at $\nu = 2/5$, $2/7$, and $2/9$, recently observed in high-quality GaAs/AlGaAs heterostructures [15–18].

The samples used were silicon MOSFET's with maximum mobility, $\mu$, around $4 \times 10^4$ cm$^2$/Vs similar to those used in Refs. [7–9,12,13]. The resistance was measured with a four-terminal dc technique using amplifiers with a very high input impedance ($> 10^{14}$ $\Omega$).

In Fig. 1 (a) we show the diagonal ($\rho_{xx}$) and Hall ($\rho_{xy}$) resistivities obtained at very low $n_s$ as functions of $1/\nu \propto B$. Despite low $n_s$, $\rho_{xy}(B)$ looks conventional, with wide plateaux at $\nu = 2$ and 1. Diagonal resistivity exhibits minima at $\nu = 6$, 2, and 1, which are also characteristic of the QHE. However, in contrast to the conventional QHE, the $\rho_{xx}$ peaks between some of these minima pierce through $\rho_{xy}(B)$ and exceed $\rho_{xy}$ by orders of magnitude.

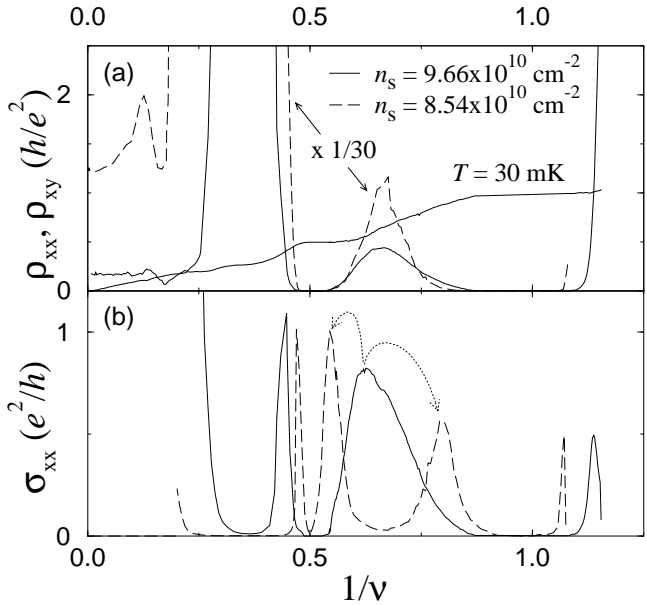

FIG. 1. $\rho_{xx}$ and $\rho_{xy}$ (a) and $\sigma_{xx}$ (b) vs $1/\nu$ for two electron densities. To avoid complications not addressed in discussion, we show $\sigma_{xx}$ only for $1/\nu > 0.2$.

The peaks with $\rho_{xx} \gg \rho_{xy}$ correspond to an insulating state with $\sigma_{xy} \ll \sigma_{xx} \ll e^2/h$. Note that lower-field $\rho_{xx}$ peaks develop directly from the QHE at $\nu = 6$ [19].

Calculating $\sigma_{xx}$ from the data for $\rho_{xx}$ and $\rho_{xy}$ (the result is shown in Fig. 1 (b)), it is possible to map the positions of extended states on the $(B, n_s)$ plane. When the Fermi energy, $E_F$, crosses an extended state, $\sigma_{xx}$ exhibits a maximum. Most of the data was obtained by sweeping gate voltage (and therefore $n_s$) at $B =$ const. We used data obtained by sweeping $B$ at constant $n_s$ only at very low $n_s$ because the phase boundary has regions almost independent of $B$ for some ranges of $n_s$.

Figure 2 shows the resulting positions of the extended states on the $(B, n_s)$ plane. Each set of symbols reflects the behavior of one $\sigma_{xx}$ peak as a function of $n_s$ and $B$, and the dotted lines represent the expected positions of the extended states at half-filled Landau levels, $\nu = i + \frac{1}{2}$. The regions between extended states correspond to the QHE with $\sigma_{xx} = 0$ and $s_{xy} \equiv \sigma_{xy} \cdot h/e^2$ shown by the numbers obtained from the data for $\sigma_{xy}$. The region with $s_{xy} = 0$, separated from regions with $s_{xy} \neq 0$ by the lowest extended state, is insulating. We will call this boundary, shown by the solid line, the "QHE/insulator (QHE/I) boundary".

The two most important features of Fig. 2 are: (i) as the magnetic field is lowered, the higher $n_s$ extended states, separating QHE with different values of $s_{xy}$, merge with the QHE/I boundary, and (ii) the QHE/I boundary oscillates as a function of magnetic field for $B \lesssim 3.5$ T reaching maxima where the higher $n_s$ extended states join. Both features are in agreement with the phase diagram obtained by SKD. They are not observed in the experimental results reported by GJJ. We note that merging of extended states with a boundary between QHE and zero-field metallic state was suggested by HLR.

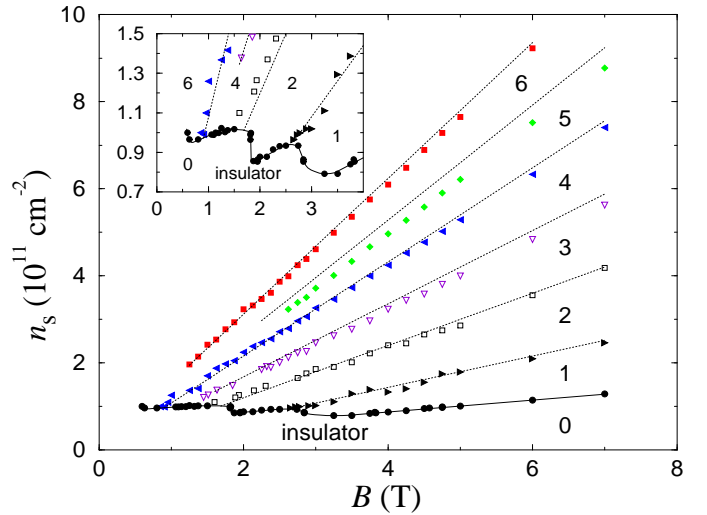

FIG. 2. The map of extended states. Numbers show $\sigma_{xy}$ in units of $e^2/h$. The inset shows detais at low $B$ and $n_s$.

Using these experimentally obtained positions of the extended states, we reconstruct a "disorder vs filling factor" phase diagram for the QHE to compare with the theoretical results of KLZ and HLR. To characterize the disorder quantitatively, we used the $B = 0$ electron scattering time extracted from $\mu(n_s)$ at $T = 1.8$ K. The resulting phase diagram is shown in Fig. 3. The level structure in silicon is complex, with both spin and valley splittings associated with each Landau level. Thus, QHE states with $s_{xy} = 1$ to 4 all belong to Landau level zero, and the merging of different extended states belonging to this level is trivial. However, the QHE state with $s_{xy} = 6$ belongs to Landau level one. Therefore, the experimentally obtained phase diagram principally differs from the theoretical one because *it allows a direct transition from insulator to QHE with $s_{xy} = 6$, associated with the next Landau level*. The QHE/I boundary (the upper solid line) oscillates reaching maxima near $\nu = 1$, 2, and 6, and minima at the points where it merges with extended states separating neighboring QHE regions [7,12,13].

We believe that these oscillations of the QHE/I boundary may be caused by many-body (MB) enhancements of energy gaps between levels which occur at integer filling factors [20]. Indeed, suppose that the "starting" phase diagram (i.e., the diagram which does not include any MB enhancement) looks like the schematic diagram in the inset in Fig. 4. The QHE/I boundary is monotonic for $1/\nu < 0.5$, and extended states, which separate QHE with different $s_{xy}$, merge with it (for simplicity, we consider spinless electrons and ignore the valley splitting; however, we assume that all gaps are subject to MB enhancement). Then, we "turn on" the MB enhancement of gaps between the neighboring levels when $E_F$ is at integer $\nu$. This enhancement results in the penetration of the QHE, associated with that $\nu$, deeper into insulating region, as shown by vertical arrows. At the same time, the points where the extended states merge with the QHE/I boundary, remain on the "starting" boundary because the enhancement is unimportant when $E_F$ coincides with an extended state. The result is an oscillating QHE/I boundary [21]. This picture leads to a reconsiliation of the results of GJJ and SKD. For the case of pure Landau level gaps at even $\nu$ in GaAs systems, like the case for GJJ, there is little MB enhancement of the gaps. Therefore, the QHE/I boundary is monotonic and a multi-reentrant insulating state is not seen.

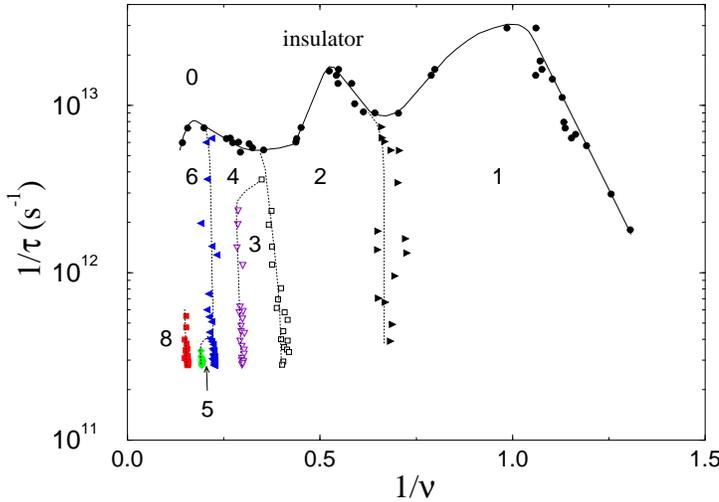

FIG. 3. Disorder *vs* filling factor phase diagram recalculated from the data shown in Fig. 2. The dotted lines show the approximate behavior of the extended states, and the solid line is the QHE/I boundary. Numbers between the lines give $\sigma_{xy}$ in units of $e^2/h$.

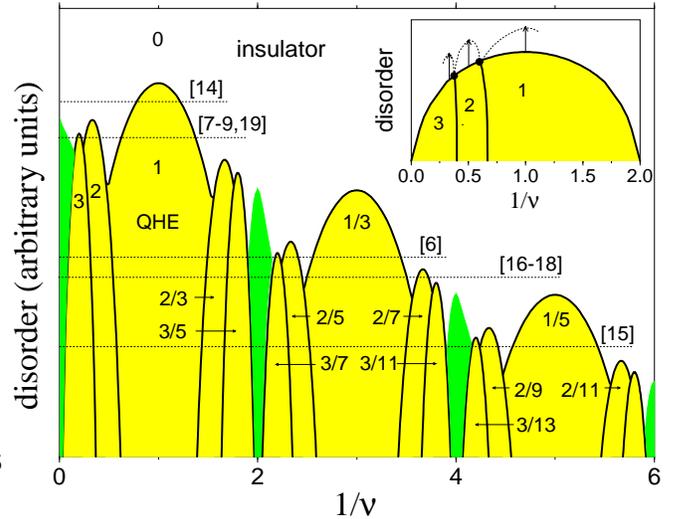

FIG. 4. A suggested phase diagram for the quantum Hall effect. Shaded regions correspond to the QHE with $\sigma_{xx} = 0$ and values of $s_{xy}$ given by numbers. Thick solid lines separating regions of QHE with different $s_{xy}$ designate the extended states. The white region corresponds to the insulator with $s_{xy} = \sigma_{xx} = 0$, and the darker shaded regions correspond to the "metallic" states near $B = 0$ [25] and half-integer filling factors. The horizontal lines trace out transitions between QHE and insulating phases consistent with the experimental references with which they are labeled. Inset illustrates a possible mechanism leading to oscillations of the QHE/I boundary, see text.

The question may arise why the QHE/I boundary shows maxima at $\nu = 1$, 2, and 6, and not at $\nu = 4$. We believe this is due to the lack of MB enhancement at $\nu = 4$ and to complex level structure in Si MOSFET's. Splittings at $\nu = 4$ and 8 vanish at $B \lesssim 1$ T, as shown by direct measurements of the chemical potential [22].

The merging of extended states separating two different QHE states with the QHE/I boundary can be seen from the bare data for $\rho_{xx}$ and $\rho_{xy}$. Consider a trivial situation between QHE minima at, *e.g.*, $\nu = 1$ and 2 (Fig. 1 (a)). As long as the $\rho_{xx}$ peak lies below $\rho_{xy}$ (lower disorder, the solid curve), its position approximately coincides with the maximum of $\sigma_{xx}$ (see Fig. 1 (b)) and, therefore, with the position of the extended state separating QHE with $s_{xy} = 1$ and 2. However, as soon as the peak of $\rho_{xx}$ pierces through $\rho_{xy}$ (higher disorder, the dashed curve), the single peak of $\sigma_{xx}$ splits into two peaks separated by an insulator, as shown in Fig. 1 (b) by arrows. In other words, the extended state between QHE with $s_{xy} = 1$ and 2 becomes a part of QHE/I boundary. A similar merging happens every time a peak of $\rho_{xx}$ overtakes $\rho_{xy}$, *e.g.*, between QHE minima with $s_{xy} = 2$ and 6 and possibly 6 and 10 [19]. More interestingly, *this happens also between some of neighboring FQHE minima, e.g.,* between $\nu = 2/5$ and $1/3$ [16,18]. This $\rho_{xx}$ peak grows by an order of magnitude as a result of decreasing hole density (see Fig. 1 in Ref. [16]) and overtakes $\rho_{xy} \approx \frac{5}{2} h/e^2$ [18]. Therefore, the extended state between fractions $2/5$ and $1/3$ merges with the QHE/I boundary in a way similar to the extended state between $s_{xy} = 1$ and 2 considered above.

The similarity between the integer QHE (IQHE) at $\nu = 1, 2, 3 \ldots$ and FQHE with $\nu$ equal, *e.g.*, to $1/3$, $2/5$, $3/7 \ldots$ follows from the composite fermion approach of Jain [23] and HLR. It can also be seen in the global phase diagram of KLZ. Moreover, the metal/insulator phase diagram for FQHE, experimentally obtained by SKB [24], looks similar to that for the IQHE. This similarity prompts us to generalize the experimentally obtained IQHE phase diagram to FQHE using correspondence principles presented by KLZ. For simplicity, we ignored spin and valley splittings. In this case, the regions with $s_{xy} = 1\ldots4$ collapse into a single region with $s_{xy} = 1$, the regions with $s_{xy} = 5\ldots8$ — into a single region with $s_{xy} = 2$ *etc.* We used two main features of the experimental phase diagram shown in Fig. 3: (i) merging of the upper extended states with QHE/I boundary, and (ii) the oscillatory QHE/I boundary. The suggested resulting phase diagram is shown schematically in Fig. 4. The main feature of this phase diagram is that it allows direct transitions between insulating states and QHE with $s_{xy} = 2, 3, 2/3, 2/5, 2/7, 2/9, 2/11$, which are forbidden according to the global phase diagram of KLZ and HLR. However, recently there have been a number of reports on experimental observation of direct transitions between insulator and QHE with $s_{xy} = 6$ and possibly 10 [19], $s_{xy} = 2/5$ [16,18], $s_{xy} = 2/7$ [17,18], and $s_{xy} = 2/9$ [15], which are consistent with the proposed phase diagram. Horizontal dotted lines show schematically the transitions realized in some of these experiments. It can be seen that in principle there should exist a direct transition between insulator and QHE with $s_{xy} = 2/11$.

In summary, we report an experimentally determined map of extended states in a magnetic field. We show that the higher-$n_s$ extended states can merge with the lowest one, in contrast to theoretical predictions. Using our data, we construct a "disorder vs filling factor" phase diagram for the IQHE and show that the generalization of our results to the FQHE explains recently observed transitions between insulating states and FQHE states with $s_{xy} = 2/5, 2/7$, and $2/9$.

We thank V. T. Dolgopolov, A. G. Golitsyna, G. V. Kravchenko, B. A. Mason, K. Mullen, S. Q. Murphy, M. B. Santos, and A. A. Shashkin for useful discussions and help with preparation of this manuscript. This work was supported by grants DMR 89-22222 and Oklahoma EPSCoR via LEPM from the National Science Foundation, a grant 94-02-04941 from Russian Foundation for Basic Research, and MUG000 from the International Science Foundation.